\documentclass[prc,twocolumn,secnumroman,tightenlines,amssymb,aps,nobibnotes,
  superscriptaddress,showpacs,balancelastpage,floatfix,nofootinbib]{revtex4}
\usepackage{graphicx,colordvi}
\usepackage{multirow}
\usepackage{color}
\usepackage[utf8]{inputenc}
\usepackage{amsmath}
\usepackage{ulem}

\begin{document}

\title{\bf Towards a unified description of isotopic fragment properties in
spontaneous and fusion-induced fission within a 4D dynamical Langevin model}
\author{K. Pomorski}\email{pomorski@kft.umcs.lublin.pl}
\affiliation{Maria Curie Sk\l odowska University, Department of Theoretical Physics,
 20031 Lublin, Poland}
\author{B. Nerlo-Pomorska}
\affiliation{Maria Curie Sk\l odowska University, Department of Theoretical Physics,
 20031 Lublin, Poland}
\author{J. Bartel}
\affiliation{IPHC/DRS and University of Strasbourg, 67200 Strasbourg, France}
\author{C. Schmitt}
\affiliation{IPHC/DRS and University of Strasbourg, 67200 Strasbourg, France}
\author{Z. G. Xiao}\email{xiaozg@mail.tsinghua.edu.cn}
\affiliation{Department of Physics, Tsinghua University, Beijing 100084, China}
\author{Y. J. Chen}
\affiliation{China Institute of Atomic Energy, Beijing 102413, China}
\author{L. L. Liu}
\affiliation{China Institute of Atomic Energy, Beijing 102413, China}

\pacs{24.75.+i, 25.85.-w,28.41.A }
\date{\today}
\begin{abstract}
\noindent

Spontaneous fission of $^{252}$Cf and fusion-induced fission of $^{250}$Cf are investigated within a multi-dimensional Langevin model. The potential-energy surface is calculated in the macroscopic-microscopic LSD+Yukawa-folded approach using the four-dimensional Fourier-over-Spheroid shape parametrization. The dynamical evolution described by the Langevin equation is coupled to neutron evaporation, thereby allowing for the possibility of multi-chance fission. Charge equilibration and excitation-energy sharing between the fragments emerging at scission are evaluated, and their de-excitation is finally computed. The correlation between various observables, particularly the isotopic properties of the fragments, is discussed and compared with the experiment whenever available. The theoretical predictions are generally in good agreement with the data.\\
\\
\noindent
KEYWORDS: nuclear fission, macro-micro model, fission fragment isotopic and TKE  yields, pre- and post-scission neutron multiplicities
\end{abstract}
\maketitle

%%%%%%%%%%%%%%%%%%%%%%%%%%%%%%%%%%%%%%%%%%%%%%%%%%%%%%%%%%%%%%%%%%%%%%%%%%%%%%
  
\section{Introduction}

The nuclear fission phenomenon, discovered in 1938, continues to be of primary interest in nuclear physics both from the fundamental and applications point of view. In this context, accurately reproducing the mass, charge, isotopic, and total kinetic energy (TKE) yields of fission fragments and the multiplicities of emitted neutrons is a stringent test of any modern theoretical model. A representative selection of contemporary models of various types developed by different groups can be found in Refs.~\cite{RMo13,MSc22,ACD21,MJV19,USANG2019, SIM2018,IVA2024,VRET2023,BUL2016,SADHU22,VER2021, REG2019,ARIT2022,LM2019, PASCA,ROD2014,PNS23}. For an overall picture of modern fission theories and perspectives, we refer to recent reviews in Refs.~\cite{BBB20, SJA16, SCHUN2022}.

The present investigation is a continuation of our previous studies \cite{PIN17,PDH20,PBK21,LCW21,KDN21} in which fragment mass yields for fission at low excitation energy were investigated in a wide range of fissioning systems from pre-actinides to trans-actinides. For some specific actinides, TKE yields were also studied \cite{LCW21}. We recently substantially extended these investigations in Refs.~\cite{PNe23,PDN23,PNS23}. In particular, a model of charge equilibration of the fragments at scission was introduced, allowing us to go beyond the widespread Unchanged-Charge-Density (UCD) assumption. In addition, the Langevin equation was coupled to a Master-type equation for modeling the possible emission of neutrons from the excited fissioning system prior to scission and from the primary fragments after scission. As for the latter, a simple prescription for sharing the excitation energy between the fragments at scission was implemented. In our most recent calculations \cite{PNS23}, the nuclear shape description is based on the so-called Fourier-over-Spheroid (FoS) parametrization, which is an innovative variant of the original Fourier shape parametrization presented in \cite{SPN17}. As discussed in Ref.~\cite{PNe23}, the FoS parameterization is better adapted to fission calculations on a large grid. It is to be emphasized that the extensions \cite{PNe23,PDN23,PNS23} of our original model are mandatory for any meaningful calculation of fragment ($A$, $Z$) isotopic yields. This new approach offers the possibility to study fission in detail, as illustrated in recent experimental campaigns \cite{camaano2015, RCF19, martin2021}.

In the present study, we use the advanced version of our model \cite{PNS23} to address the fission of two californium isotopes in two excitation-energy regimes. In particular, we consider spontaneous fission of $^{252}$Cf, and fission of $^{250}$Cf at an excitation energy $E^*$ of 46 MeV induced by the fusion reaction generated by a $^{238}$U beam on a $^{12}$C target. Experimental isotopic yields for both systems are available from Refs.~\cite{ATJ20,COH06,PCD20,MDA20,WAS23} and Refs.~\cite{CDF13,RCF19}, respectively. Comparison with these data allows us to evaluate our theoretical model's performance over a wide range of excitation energies (our previous studies have focused on low-energy fission). Such a study will allow for a strict test of the assumed evolution of various quantities with temperature.
 
The main features of the model, which are important for an understanding of the present study, are briefly recalled in Section II, while we refer to Refs. \cite{KDN21,PNS23} for further details and parameters. Sections III and IV present the calculated results for spontaneous fission of $^{252}$Cf and fusion-induced fission of $^{250}$Cf at excitation energy $E^*$ = 46 MeV. Summary and concluding remarks are given in Section V. 
                          
%%%%%%%%%%%%%%%%%%%%%%%%%%%%%%%%%%%%%%%%%%%%%%%%%%%%%%%%%%%%%%%%%%%%%%%%%%%%%%
\section{Model}

\subsection{Shape parametrization and the potential-energy surfaces}
\label{Sec2.1}

The model used in our present study is the same as in our previous investigation \cite{PNS23} on thermal neutron-induced fission of $^{235}$U. That is why only its main ingredients are shortly listed below. Using what we call the Fourier-over-Spheroid shape parametrization developed in Ref.~\cite{PNe23}, the surface of a deformed nucleus is described in cylindrical coordinates $(\rho,\varphi,z)$ by the following formula: 
\begin{equation}
\rho_s^2(z,\varphi)=\frac{R_0^2}{c}\,f\left(\frac{z-z_{\rm sh}}{z_0}\right)
{1-\eta^2\over 1+\eta^2+2\eta\cos(2\varphi)} ~.
\label{rhos}
\end{equation}
Here $\rho_s(z,\varphi)$ is the distance of a surface point to the $z$-axis. The function $f(u)$ defines the shape of the nucleus having half-length $c=1$:
\begin{equation}
  f(u)=1-u^2-\sum\limits_{k=1}^n \left\{a_{2k}\cos({k-1\over 2}\pi u)
             +a_{2k+1}\sin(k\pi u)\right\}~,
\label{fos}
\end{equation}
with
\begin{equation}
  u = \frac{z-z_{\rm sh}}{z_0}~,
\label{udef}
\end{equation}
where $z_0=c R_0$, with $R_0$ being the radius of the sphere, is the half-length of the deformed nucleus and the shift parameter  $z_{\rm sh} = -3/(4\pi)\,z_0(a_3-a_5/2+\dots)$ ensures that the origin of the coordinate system is located at the center of mass of the nucleus so that $-1 \leq u \leq 1$. The expansion coefficients $a_i$ are treated as the deformation parameters. The first two terms in $f(u)$ describe a sphere. The volume conservation condition implies $a_2=a_4/3-a_6/5+\dots$. The parameter $c$ determines the elongation of the nucleus, keeping its volume fixed, while $a_3$ and $a_4$ are, respectively, the deformation parameters essentially responsible for the reflection asymmetry and the neck formation of the deformed shape. The parameter $\eta$ in Eq.~(\ref{rhos}) allows for a possible non-axial deformation of the nucleus. 

Equation (\ref{fos}) is entirely equivalent to the one based on the original Fourier expansion of Ref.~\cite{SPN17} but is easier to handle in the case of fission because, in the present case, and contrary to the original definition, the range of variability of the $a_i$ coefficients does not depend on the elongation $c$. In addition, the mass ratio of the fragments, their relative distance, and the radius of the neck between them, measured in $z_0$ units, do not depend on the elongation of the nucleus. In addition, the heavy fragment mass-number $A_{\rm h}$ is nearly a linear function of the $a_3$ deformation: $A_{\rm h}\approx (1+a_3){A\over 2}$ at the scission configuration ($a_4\approx 0.72$).  One has also to note that for the reflection-symmetric shapes ($a_3=0$), the geometrical scission point occurs when $a_4=a_4^{\rm sc}={3\over 4}+{6\over 5}a_6\dots$ independently of the elongation $c$.

The potential energy surfaces (PES) of fissioning nuclei are then obtained in the 4D space of deformation parameters ($c,a_3,a_4,\eta$) using the macroscopic-microscopic (macro-micro) model \cite{NTS69}. The macroscopic part of the energy is evaluated according to the Lublin-Strasbourg-Drop (LSD) formula \cite{PDu09}, while the microscopic energy corrections are calculated using the Yukawa-folded single-particle potential \cite{DPB16} and the Strutinsky shell correction method \cite{Str66,NTS69}. The pairing correlations are described using the BCS formalism with  an approximative projection on good particle number \cite{GPo86,PPS89}. All parameters of the macro-micro model used in the present study are the same as in Ref.~\cite{PDN22}. 

\begin{figure}[htb]
\includegraphics[width=0.45\textwidth]{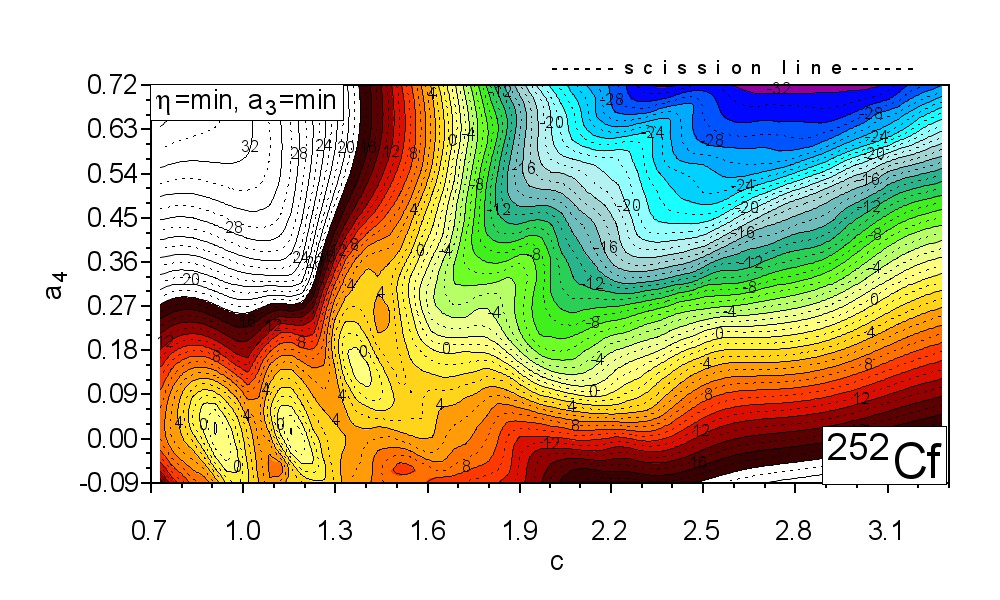}\\[-2ex]
\includegraphics[width=0.45\textwidth]{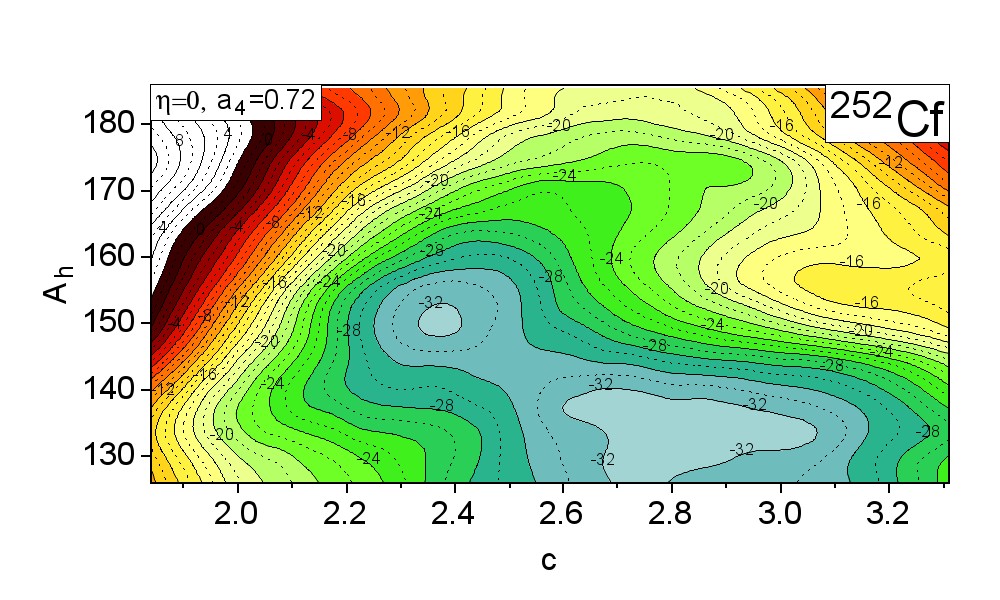}\\[-4ex]
\caption{Potential energy surface of $^{252}$Cf projected onto the ($c,\,a_4$) plane (top) and its ($c,A_{\rm h}$) cross-section (bottom) around the scission configuration at $a_4=0.72$, where A$_{\rm h}$ is the mass of the heavy fragment. Each point of the top map is minimized concerning the non-axial ($\eta$) and the $a_3$ shape variables, respectively. The values of the energy layers are taken relative to the spherical liquid drop binding energy.}
\label{Cf252pes}
\end{figure}

Please recall here that due to energy-dissipation effects, even spontaneously fissioning nuclei get excited near the scission configuration. The resulting temperature effect of atomic nuclei is even more crucial in the case of neutron-induced fission or the fission of compound nuclei formed in heavy-ion collisions. It would not be easy to evaluate the PES with changing temperature $T$ on the way to the scission configuration. Therefore,  we do it approximately in the following way. In the macro-micro model, one generally assumes that the total potential energy
\begin{equation}
V_{\rm tot}=V_{\rm mac}+ V_{\rm mic}
\label{macmic}
\end{equation}
is the sum of the macroscopic $V_{\rm mac}$ and microscopic $V_{\rm mic}$ parts. The macroscopic part of the potential energy grows parabolically with increasing temperature (refer to, e.g., Ref.~\cite{NPB02}), while the amplitude of the microscopic energy correction decreases. Following the estimates made in Ref.~\cite{NPB06} we have assumed that the microscopic part of the potential energy varies with temperature $T$ according to the following phenomenological relation \cite{KDN21}:
\begin{equation}
 V_{\rm mic}(\vec q,T)\approx {V_{\rm mic}(\vec q,T=0)\over 1+\exp((T-1.5)/0.3)}~,
\label{EmicT}
\end{equation}
where the temperature $T$ is in MeV units and $\vec q$ stands for the $\{c,a_3,a_4,\eta\}$ deformation.

%%%%%%%%%%%%%%%%%%%%%%%%%%%%%%%%%%%%%%%%%%%%%%%%%%%%%%%%%%%%%%%%%%%%%%%%%%%%%%%%

\subsection{Dynamical evolution}
\label{Sec2.2}

In our approach, the dissipative fission dynamics is described by the Langevin equation. In the generalized coordinates ($\{q_i\},~~i=1,2,...,n$) it has the following form \cite{KPo12}:
\begin{equation}
\begin{array}{ll}
{dq_i\over dt} =& \sum\limits_{j} \; [{\cal M}^{-1}(\vec q\,)]_{i\, j} \; p_j  \\
 {dp_i\over dt}=& - {1\over 2} \sum\limits_{j,k} \,
           {\partial[{\cal M}^{-1}]_{jk}\over\partial q_i}\; p_j \; p_k
          -{\partial V(\vec q)\over\partial q_i}\\
          &- \sum\limits_{j,k} \gamma_{ij}(\vec q) \;
           [{\cal M}^{-1}]_{jk} \; p_k + F_i(t) \,\,, 
\end{array}
\label{LGV}
\end{equation}
Here $V(\vec q\,)=E_{\rm pot}(\vec q\,)-a(\vec q\,)T^2$ is the Helmholtz free-energy of the fissioning nucleus with temperature $T$ and $a(\vec q\,)$ is the single-particle level density parameter. The potential energy $E_{\rm pot}(\vec q\,)$ at a given deformation $\vec q$ is obtained by the macro-micro prescription as stated above. The parameter $a(\vec q\,)$ is, according to Ref.~\cite{NPB02}, a deformation-depending function. The inertia and friction tensors ${\cal M}_{jk}$ and $\gamma_{ij}$ are respectively evaluated in the irrotational flow and the wall approximation, as described in Refs.~\cite{BNP19,KDN21}.

The vector $\vec F(t)$ stands for the random Langevin force, which couples the collective dynamics to the intrinsic degrees of freedom and is defined as:
\begin{equation}
F_i(t) \!\!=\!\! \sum_{j} g_{ij}(\vec q\,) \; G_j(t) \,\,,
\label{rforce}
\end{equation}
where $\vec G(t)$ is a stochastic function whose strength $g(\vec q\,)$ is given by the diffusion tensor ${\cal D}(\vec q\,)$ defined by the generalized Einstein relation:
\begin{equation}
{\cal D}_{ij} \!\!=\!\!T^*\gamma_{ij} \!\!=\!\! \sum_{k} g_{ik} \; g_{jk}~,
\label{Eirel}
\end{equation}
with
\begin{equation}
T^*=E_0/{\rm tanh}\left({E_0\over T}\right)~,
\label{Tstar}
\end{equation}
The vector function $\vec G(t)$ takes into account both statistical and collective fluctuations \cite{PHo81}. In the following, we have taken $E_0=3\times 0.5$ MeV, assuming that each collective mode contributes 0.5$\,$MeV to the zero-point energy. 
The temperature $T$ is obtained from the thermal excitation energy $E^*$ defined as the difference between the initial energy $E_{\rm init}$ and the final energy, which is the sum of kinetic ($E_{\rm kin}$) and potential ($V$) energies of the fissioning nucleus at the present deformation ($\vec q$) and the sum of the binding and the kinetic energies of emitted particles ($E_{\rm part}$)
\\[-2ex] 
\begin{equation}
 a(\vec q\,)T^2=E^*(\vec q\,)=E_{\rm init}-[E_{\rm kin}(\vec q\,)+V(\vec q\,)
 +E_{\rm part}]~.
\label{temp}
\end{equation}
The initial conditions of the dynamical calculation correspond to the excited compound system in the vicinity of the outer saddle point, e.g., for $^{252}$Cf: $c\approx 1.6,\,a_3\approx 0.15,\,a_4\approx 0.12,\,\eta=0$.  We assume that scission takes place when the neck parameter $a_4$ is equal to 0.72 since this value corresponds to a neck radius approximately equal to the nucleon radius $r_{\rm neck}\!=\!r_0=1,217\,$fm. Non-axiality was found to be significant only at small elongations before reaching the outer saddle ($c \approx 1.6$ for the systems considered here), consistent with what had been found in the past within various approaches \cite{PBK21}. At larger deformations, its influence is negligible. Moreover, the role of higher-order Fourier expansion coefficients $a_5$ and $a_6$ in Eq.\ (\ref{fos}) is small even in the region of well-separated fission fragments, as shown in Ref.~\cite{KDN21}. Consequently, we restrict the Langevin calculations to the 3D ($c,\, a_3,\, a_4$) deformation space when discussing fission dynamics.

Using the above formalism and procedure, we have performed extended dynamical calculations, including around $10^5$ fissioning Langevin trajectories, from which we extracted the predictions of the model for various observables such as the fission fragment mass, charge, or kinetic energy distributions. Please note that we have used the same set of parameters as the one employed in our previous study \cite{PNS23} in which neutron-induced fission of $^{235}$U and bimodal fission of Fermium isotopes were discussed.

The mass of the heavy ($A_{\rm h},\,\vec q_{\rm h}$) and the light fragments ($A_l,\,\vec q_l$) are proportional to the volumes of the daughter nuclei at the scission point, which defines the end of each Langevin trajectory. 

Knowing the fragment deformations at scission $\vec q_l$ and $\vec q_{\rm h}$, it is possible to find the most probable charge for each isobar by analyzing the energy of the system at scission as a function of the charge number $Z_{\rm h}$ of the heavy fragment:
\begin{equation}
\begin{array}{rl} 
E(Z_{\rm h};&Z,A,A_{\rm h},\vec q_{\rm h},\vec q_l)
    =E_{\rm LSD}(Z-Z_{\rm h},A-A_{\rm h});\vec q_l)\\[+1ex]
    +&E_{\rm LSD}(Z_{\rm h},A_{\rm h};\vec q_{\rm h})
     +E_{\rm Coul}^{\rm rep}-E_{\rm LSD}(Z,A;0)~,
\end{array}
\label{echeq}
\end{equation}
where $A_{\rm h}$ is the heavy fragment mass number and the fragment Coulomb repulsion energy $E_{\rm Coul}^{\rm rep}$ is given by
\begin{equation}
\begin{array}{ll}
E_{\rm Coul}^{\rm rep}&={3e^2\over 5r_0}\left[{Z^2\over A^{1/3}}B_{\rm Coul}(\vec q_{\rm sc})\right.\\[1ex] 
 &\left.-{Z_{\rm h}^2\over A_{\rm h}^{1/3}}\,B_{\rm Coul}(\vec q_{\rm h}) 
  -{Z_l^2\over A_l^{1/3}}\,B_{\rm Coul}(\vec q_l)\right]\,.
\end{array}
\label{ECrep}
\end{equation}
Here, $r_0=1.217$ fm and the Coulomb shape function $B_{\rm Coul}$ is the same as in the LSD mass formula \cite{PDu09}.

The distribution of the heavy-fragment charge number can be estimated using a Wigner function corresponding to the energy $E$ obtained with the help of Eq.~(\ref{echeq}) for different values of $Z_{\rm h}$ (refer to Ref.~\cite{PNS23} for more details):
\begin{equation}
 W(Z_{\rm h})=\exp\{-[E(Z_{\rm h})-E_{\rm min}]^2/E_{\rm W}^2]~.
\label{Wigner}
\end{equation}
This function gives the probability distribution of the fragment charge. The energy $E_{\rm min}$ in Eq.~(\ref{Wigner}) is the lowest discrete energy in (\ref{echeq}) as a function of $Z_{\rm h}$. Furthermore, a random number \cite{PNS23} is introduced to determine the charge number $Z_{\rm h}$ of the heavy fragment, while the charge number of the light fragment is $Z_l=Z-Z_{\rm h}$. The energy $E_{\rm W}$ should be chosen comparable with the energy distance $\hbar\omega_0$ between harmonic oscillator shells since we are dealing here with a single-particle (proton-neutron) transfer between the touching fragments due to the charge equilibration. In the following we have assumed $E_{\rm W}=0.5\,\hbar\omega_0$. The above charge equilibration effect must be considered at the end of each Langevin trajectory when one fixes the fission fragments' integer mass and charge numbers.

The fission fragment TKE is given by a sum of the Coulomb repulsion energy (\ref{ECrep}) of the  fragments and their the pre-fission kinetic energy ($E_{\rm kin}^{\rm rel}$) of relative motion:
\begin{equation}
{\rm TKE}= E_{\rm Coul}^{\rm rep} + E_{\rm kin}^{\rm rel}~,
\label{TKE}
\end{equation}  
This expression gives, without any doubt, a more accurate estimate of the fission-fragment kinetic energy than the frequently used point-charge approximation: TKE=$e^2 Z_{\rm h}Z_l/R_{12}$, where $R_{12}$ is the distance between the fragment mass-centers.
\begin{figure}[b!]
\includegraphics[width=0.5\textwidth]{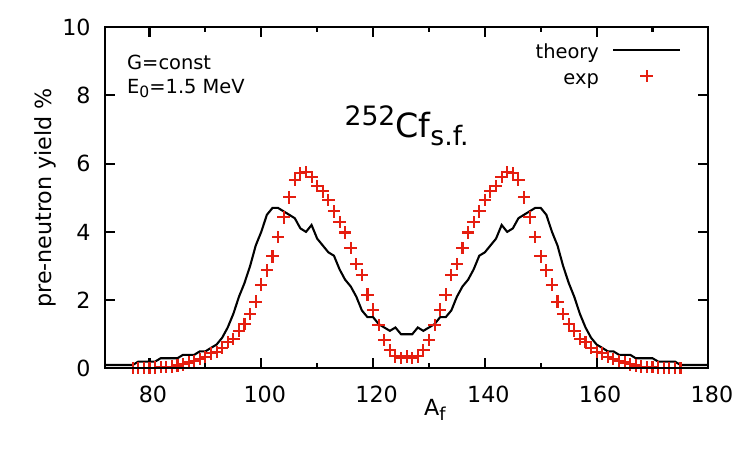}\\[-4ex]
\caption{Primary fission fragment mass yield of $^{252}$Cf as a function of mass number. The experimental data are taken from Ref.~\cite{ATJ20}.}
\label{Cf252mexp}
\end{figure}

\subsection{Neutron evaporation}
\label{Sec.2.3}

Thermally excited heavy nuclei de-excite by emitting light particles, like neutrons,  protons, or $\alpha$-particles. At relatively low excitation energies ($E^* < 80$ MeV), only neutron evaporation takes place, while the emission of a proton or $\alpha$-particle is unlikely \cite{PNS00}. Emission of high-energy $\gamma$-rays in competition with neutron evaporation is rare and is therefore neglected in the present study. At the end of the de-excitation chain, below the neutron separation energy, the remaining excitation energy and angular momentum are exhausted by the low-energy $\gamma$-rays emission. The latter stage of the decay process is not included in the model, since it does not affect the observables of interest in this work.

The modeling of neutron emission from the excited compound system along its way to scission is taken from a Weisskopf-like model described in Refs.~\cite{SDP91,PNS00}. The prescription for the de-excitation process of the excited fragments emerging at scission (hereafter called the {\it primary} fragments) has been described in detail in Sec. II-D of Ref.~\cite{PNS23} and is therefore not repeated here.

%%%%%%%%%%%%%%%%%%%%%%%%%%%%%%%%%%%%%%%%%%%%%%%%%%%%%%%%%%%%%%%%%%%%%%%%%%%%%
%%%%%%%%%%%%%%%%%%%%%%%%%%%%%%%%%%%%%%%%%%%%%%%%%%%%%%%%%%%%%%%%%%%%%%%%%%%%%%%
%
\begin{figure}[thb] 
\includegraphics[width=0.5\textwidth]{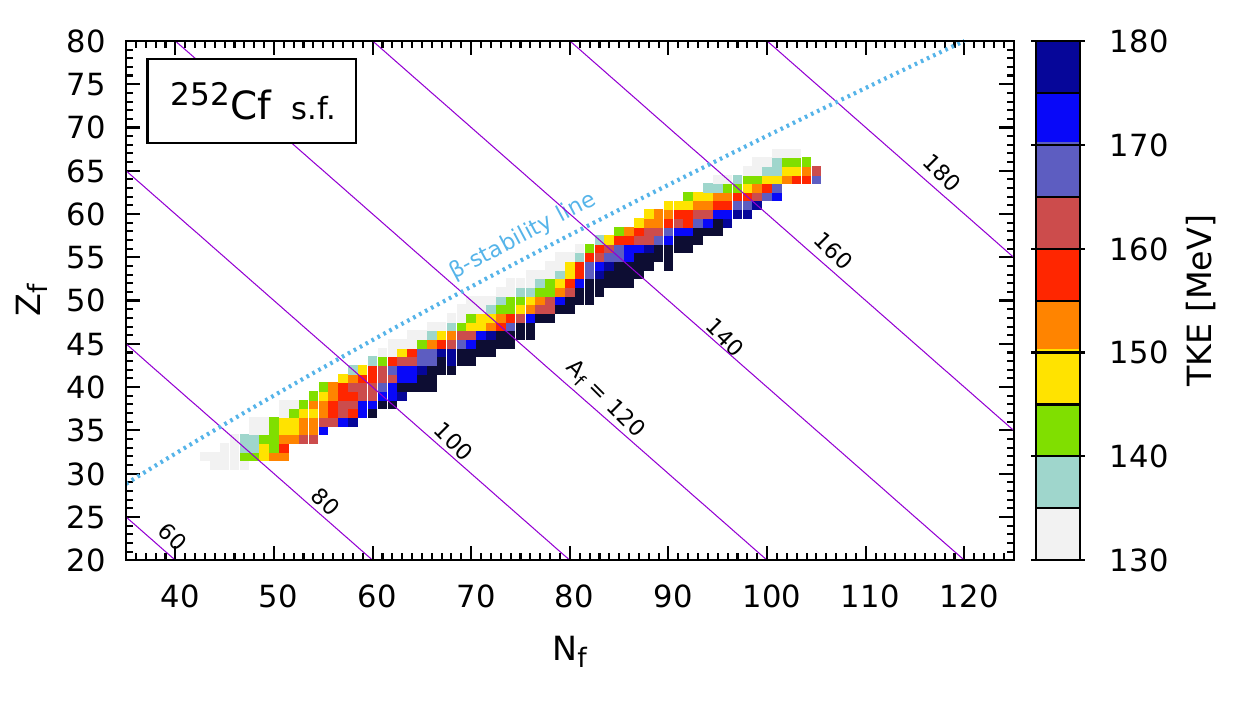}\\[-4ex]
\caption{Fission fragment total kinetic energy (TKE) of spontaneously fissioning $^{252}$Cf as a function of neutron (${\rm N_{\rm f}}$) and proton  (${\rm Z_{\rm f}}$) numbers of the primary fragments.}  %primary ???
\label{Cf252TKE}
\end{figure}

\section{Spontaneous fission yields of $^{252}$Cf}

The 4D PES of the $^{252}$Cf spontaneously fissioning nucleus is evaluated within the macro-micro model, as described in the previous section. The ($c,\,a_4$) and ($c,{\rm A_{\rm h}}$) cross-sections of the PES of $^{252}$Cf after suited minimization are presented in Fig.~\ref{Cf252pes}. The top panel shows the PES projection onto the $(c,a_4)$ plane, i.e., each energy point in the $(c,a_4)$ map is obtained by a minimization concerning the non-axial and reflection asymmetry deformation parameters $\eta$ and $a_3$ respectively. The ground state minimum (g.s.) is found at an elongation $c=1.14$ and $a_4=0.01$, while the exit point (after tunneling the fission barrier) found at $c\approx 1.6$ and $a_4\approx 0.2$, is marked by a red point. The asymmetric fission valley ends at an elongation $c\approx 2.2$ and the symmetric one at $c\approx 2.8$. The PES projection shown in the bottom panel corresponds roughly to the scission point ($r_{\rm neck}\simeq r_{\rm n}$), as noted above. From both cross sections, it can be deduced that the close-to-scission configuration of the asymmetric valley corresponds to the minimum at ${\rm A_{\rm h}} \approx 150$ and $c=2.2$. In comparison, the end of the symmetric valley is found at ${\rm A_{\rm h}}\approx 126$ and $c=2.8$. As expected, asymmetric fission of $^{252}$Cf leads to a more compact scission configuration than the more elongated one found for a symmetric splitting.
\begin{figure}[thb]
\includegraphics[width=0.45\textwidth]{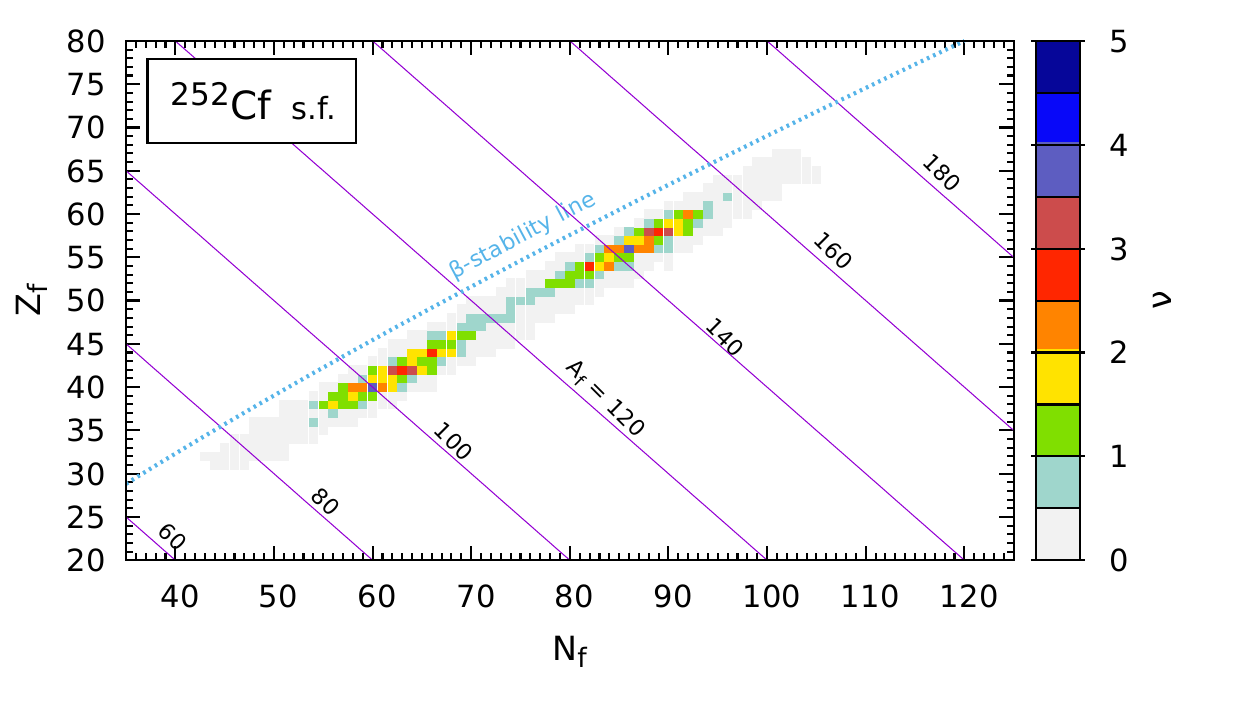}\\[-4ex]
\caption{Multiplicity of neutrons ($\nu$) emitted by each fission fragment of $^{252}$Cf as a function of neutron (${\rm N_{\rm f}}$) and proton 
(${\rm Z_{\rm f}}$) numbers of the primary fragments.}
\label{Cf252nmul}
\end{figure}

The primary fission fragment mass yield obtained in our model is compared in Fig.~\ref{Cf252mexp} with the experimental data from Ref.~\cite{ATJ20}. The theoretical yields are found to be shifted by a few mass units concerning the data. Additionally, the probability of symmetric fission is slightly overestimated.

The TKE averaged over all trajectories for each specific fragment pair is shown in Fig.~\ref{Cf252TKE} as a function of the primary fragment neutron and proton
numbers. It is seen that the neutron-rich isotopes have, in general, larger
TKEs, which means that they correspond to smaller elongations of the fissioning
system in the scission configuration. A similar map but for the multiplicity of
the neutrons emitted by the fragments is presented in Fig.~\ref{Cf252nmul}. It
is found that the symmetric fragments emit, on average, less than one neutron,
while the most probable mass asymmetric fragments evaporate around three
neutrons or more. All fission fragments are predicted in our approach to be
located below the $\beta$-stability line marked in Fig.~\ref{Cf252TKE} and thus
correspond to relatively neutron-rich isotopes, as known for fission.
\begin{figure*}[b!]
\includegraphics[width=\textwidth]{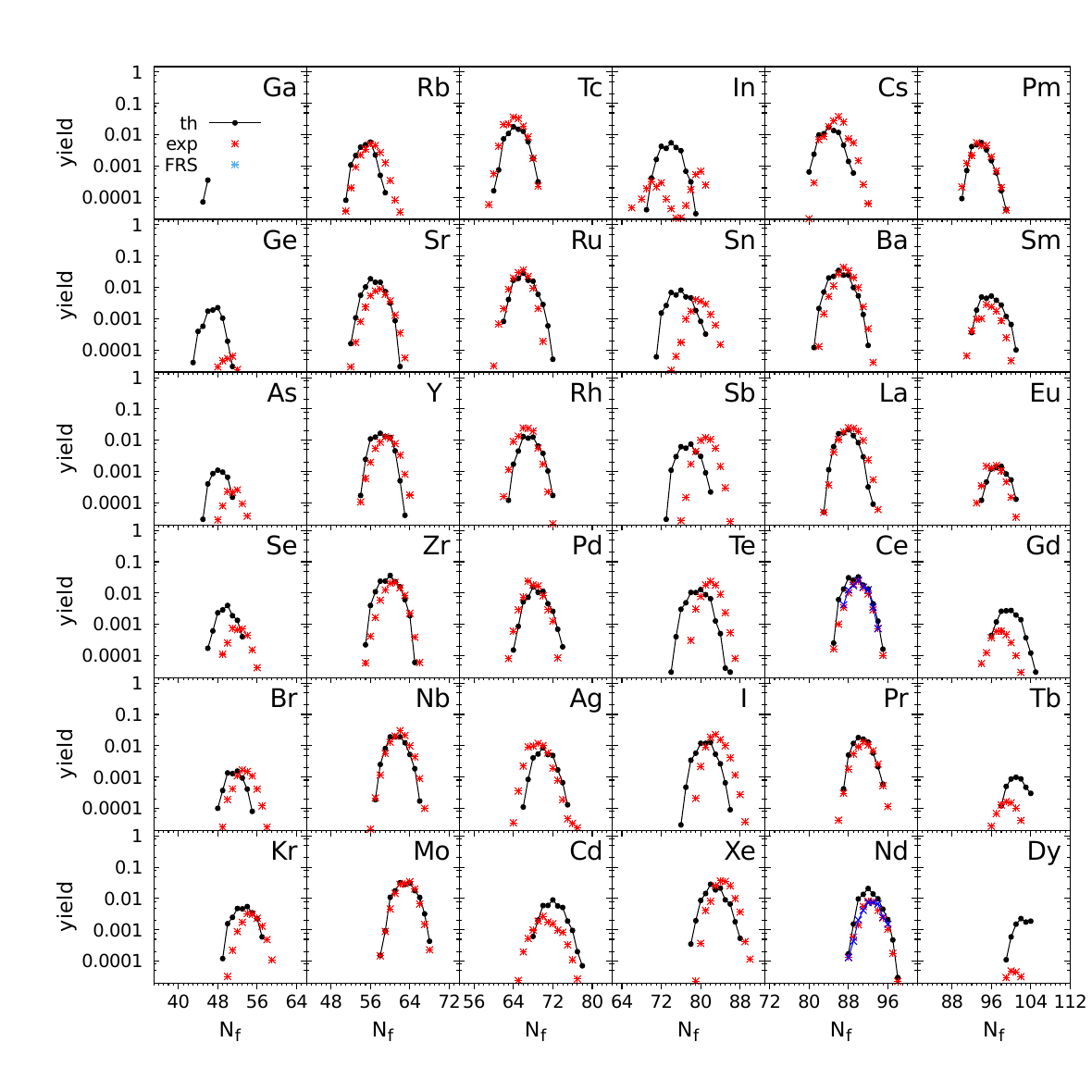}\\[-4ex]
\caption{Secondary fragment isotopic yields from Ga to Dy for spontaneous
fission of $^{252}$Cf. Black points present theoretical estimates,
while the experimental data (red stars) are taken from Refs.~\cite{COH06} and
\cite{PCD20} (for In isotopes), and for Ce and Nd from Ref.~\cite{WAS23} (blue
crosses).} \label{Cf252imy}
\end{figure*}

The calculated (black points in Fig.~\ref{Cf252imy}) secondary (i.e., after neutron evaporation) 
fragment isotopic yields are compared from Ga to Dy with the
data (red stars) taken from the ENDF/B-VII.0 and JEFF-3.3 libraries Refs.~\cite{COH06,PCD20}
and the very recent data (blue crosses) from Ref.~\cite{WAS23} via mass
measurement at the FRS ion catcher for a large range of neutron numbers.
The overall agreement of our estimates with the data is quite satisfactory,
especially when one considers that none of the model parameters was fitted to
these data. The fission yields predicted for the Sn, Te, and Xe isotopes are
slightly underestimated as compared with the experimental data of
Ref.~\cite{COH06}.

%%%%%%%%%%%%%%%%%%%%%%%%%%%%%%%%%%%%%%%%%%%%%%%%%%%%%%%%%%%%%%%%%%%%%%%%%%%%%%%

\section{Fission yields of $^{250}$Cf at $E^*$=46 MeV}

\begin{figure}[t!]
\includegraphics[width=0.45\textwidth]{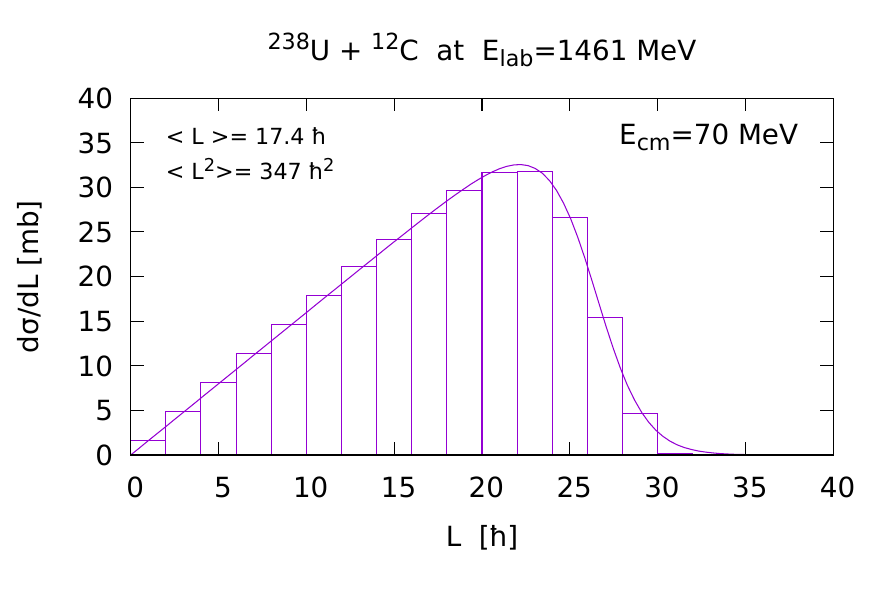}\\[-1ex]
\caption{Langevin estimate of the differential fusion cross-section of $^{250}$Cf produced in $^{238}$U (1461 MeV) + $^{12}$C collisions within a model described in Ref.~\cite{PPo94}.}
\label{Cf250fus}
\end{figure}

The development of the model of Ref.~\cite{PNS23} was initially motivated by
the wealth of experimental data available for low-energy fission, and the
importance of this energy regime in various applications. However, the energy
dependence of the model transport parameters was included already in
Ref.~\cite{PNS23}, as well as the possibility of pre-scission evaporation, {\it
i.e.} multi-chance fission. As noticed above, the model was tested only for low-energy fission inour previous work. In the present section, we
extend its application to fission at high excitation energy. Such an
investigation may serve as a stringent test for the temperature dependence of
the microscopic energy correction and the transport parameters like inertia,
friction, and diffusion tensors.

The fission fragment mass, charge, and isotopic yields for $^{250}$Cf produced
at $E^*\approx$ 46 MeV in $^{238}$U + $^{12}$C collisions were studied
experimentally in detail in Refs.~\cite{CDF13, RCF19}. The most probable angular
momentum of $^{250}$Cf is found to be around $L=20 \hbar$ as one can see in
Fig.~\ref{Cf250fus} in which the theoretical estimate of the fusion
cross-section obtained within a Langevin-type calculation  \cite{PPo94} is
presented as a function of $L$.
\begin{figure}[htb]
\includegraphics[width=0.45\textwidth]{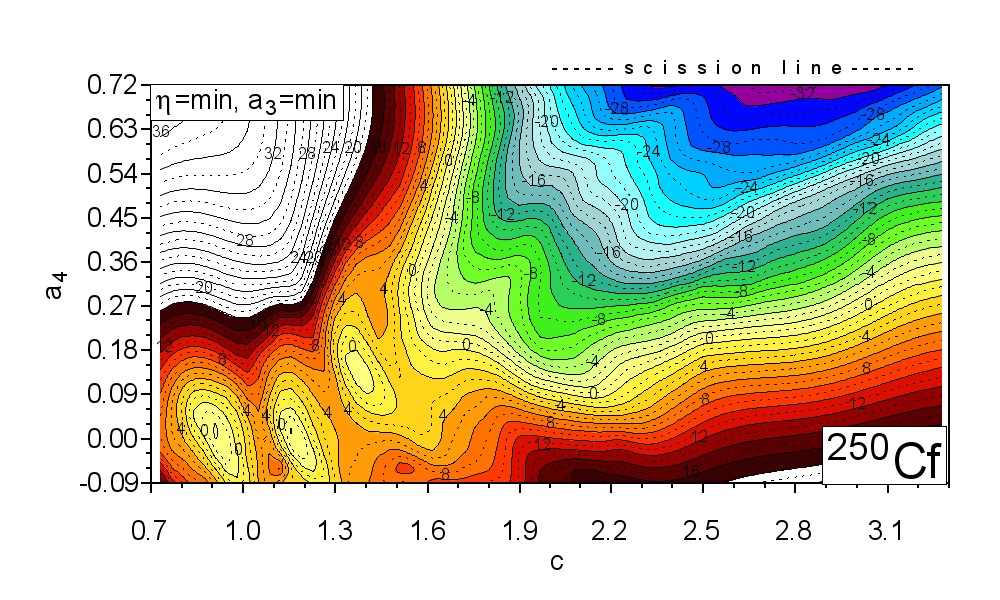}\\[-2ex]
\includegraphics[width=0.45\textwidth]{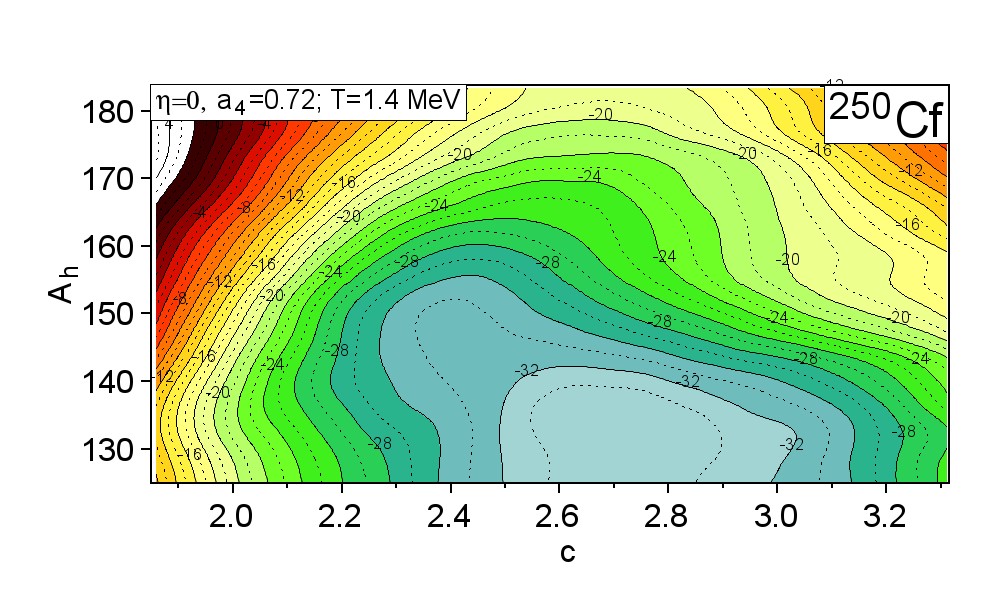}\\[-4ex]
\caption{Same as Fig.\ \ref{Cf252pes} but for $^{252}$Cf at a temperature T=1.4 MeV.}
\label{Cf250pes24}
\end{figure}

The excitation energy in the experiment above corresponds to a temperature of the $^{250}$Cf nucleus of around $T\approx 1.4$ MeV. Consequently, the amplitude of the microscopic energy corrections becomes much smaller (see Eq.~(\ref{EmicT})) than in the ground state \cite{NPB02}. The two cross sections of the PES of $^{250}$Cf evaluated for $T=1.4\,$MeV are given in the top and bottom parts of Fig.~\ref{Cf250pes24}. As anticipated, the landscapes are smoother relative to the ones at the ground state (compare, e.g., to the close-by $^{252}$Cf of Fig.~\ref{Cf252pes}) due to the shell corrections found smaller at finite temperature. Interestingly, they are, however, not fully damped. Some asymmetric fission contribution may thus persist, more or less {\it hidden} by the dominant symmetric fission component, as one can learn from the cross-section of the bottom part of Fig.~\ref{Cf250pes24}. 
\begin{figure}[htb]
\includegraphics[width=0.45\textwidth]{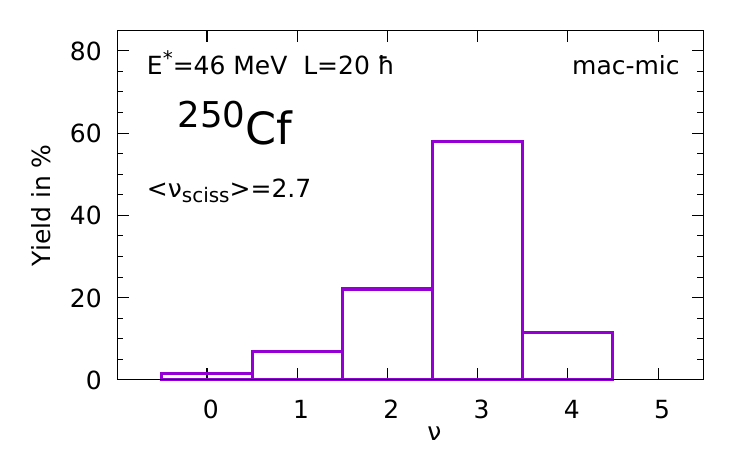}\\
\includegraphics[width=0.45\textwidth]{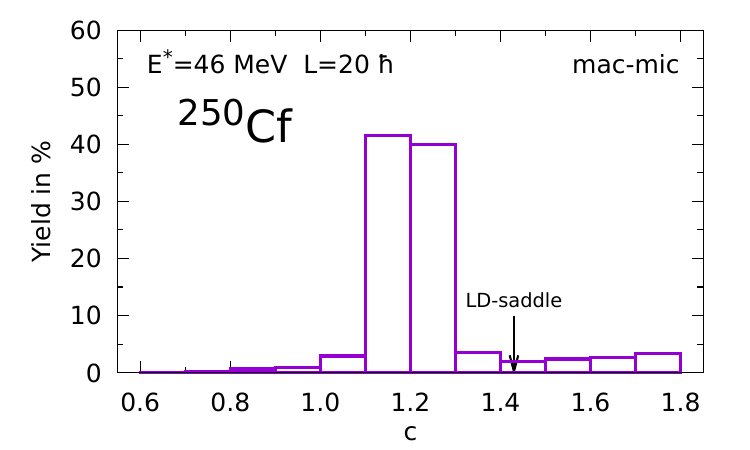}\\[-4ex]
\caption{Probability distribution of the number of neutrons emitted by $^{250}$Cf at $E^*$=46 MeV (top) and the same yield as a function of the nucleus elongation ($c$) at the moment of emission (bottom). The distributions are obtained within the Langevin+Masters model described Ref.\cite{PPo94}.}
\label{Cf250steps}
\end{figure}

Due to its relatively high initial excitation energy, the compound nucleus $^{250}$Cf produced in a fusion reaction has a high probability of emitting some neutrons before reaching the scission configuration (emission of light-charged particles prior to scission is extremely rare due to the higher energy cost \cite{PNS00}). Particle evaporation before scission leads to what is commonly called multi-chance fission. The competition between fission and evaporation is described with a set of coupled Langevin plus Masters equations, similarly to what has been done in Ref.~\cite{PNS00}, but now with the new, better adapted FoS parameterization. The yield of the number of pre-scission neutrons is presented in Fig.~\ref{Cf250steps} (top), as well as the elongation of the nucleus at which this emission takes place (bottom). One notices that most neutrons are emitted even before reaching the saddle point. The temperature of the compound nucleus decreases obviously after each emission act, so the temperature dependence of the microscopic energy (\ref{EmicT}) must be considered in our calculation. The average multiplicity of neutrons emitted before scission is found to be $\nu_{\rm pre}=2.7$, while the multiplicity of the neutrons emitted before reaching the saddle point is 2.4. One finds that the most probable (57.\%) is the emission of 3 neutrons, while the probability of events with no neutron emission, i.e., fission of $^{250}$Cf, is minimal (1.6\%).

From this result, we conclude that in the case of $^{250}$Cf at a thermal excitation energy of $E^*=46$ MeV, one is instead dealing with the fission of lighter Cf isotopes, which are, of course, due to the energy loss through the neutron emission, less excited, as one can see in Table I.
\begin{center}
\begin{table}
\caption{Distribution probability of the fissioning Cf isotopes obtained after pre-fission neutron emission and their excitation energy. $E^{\rm th}$ refers to the thermal excitation energy, i.e., after subtraction of the rotational energy.\\ }
%\begin{equation}
\begin{tabular}{|c|c|c|c|c|c|}
\hline
  $\nu_{\rm pre}$  &  4   &  3   &  2   &  1   &  0  \\[0.5ex]
\hline
      Cf           & 246  & 247  & 248  & 249  & 250 \\[0.5ex]
\hline
  yield in \%      & 11.5 & 57.9 & 22.1 &  6.9 & 1.6 \\[0.5ex]
\hline
$E^{\rm th}/{\rm MeV}$ & 15.8 & 20.4 & 27.3 & 35.7 & 45.5 \\[0.5ex]
\hline 
\end{tabular}
\label{Cf250tab}
%\end{equation}
\end{table}
\end{center}

\begin{figure}[htb]
\includegraphics[width=0.45\textwidth]{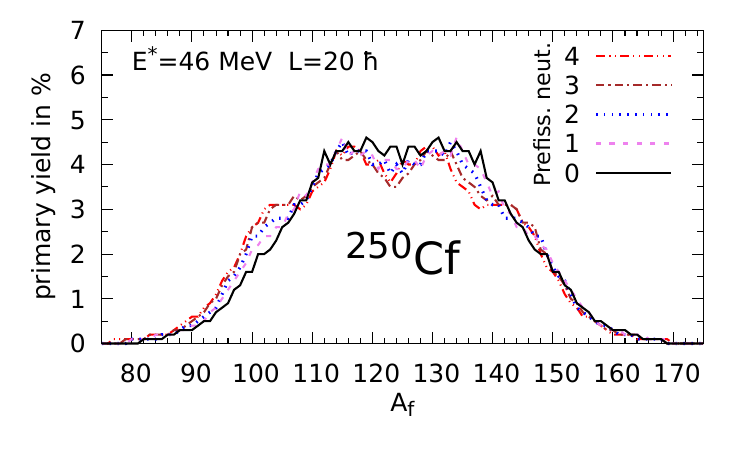}\\[-2ex]
\includegraphics[width=0.45\textwidth]{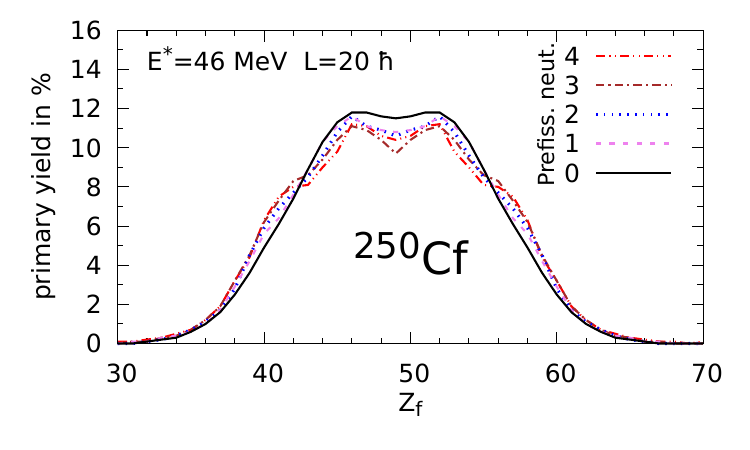}\\[-2ex]
\includegraphics[width=0.45\textwidth]{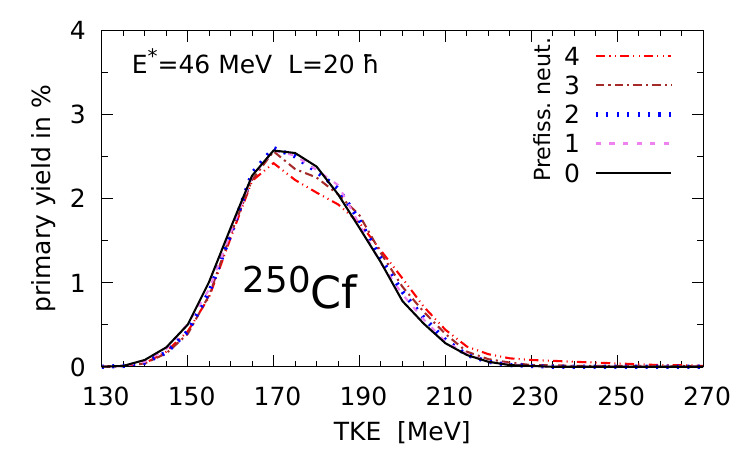}
\caption{Primary (i.e., at scission configuration) fission fragment mass (top),
charge (middle), and TKE (bottom) yields corresponding to the different fission
chances.}
\label{Cf250yn}
\end{figure}

\begin{figure}[htb]
\includegraphics[width=0.45\textwidth]{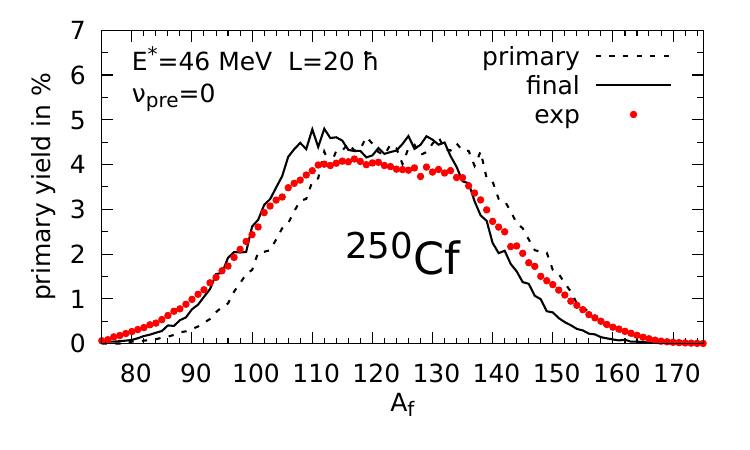}\\
\includegraphics[width=0.45\textwidth]{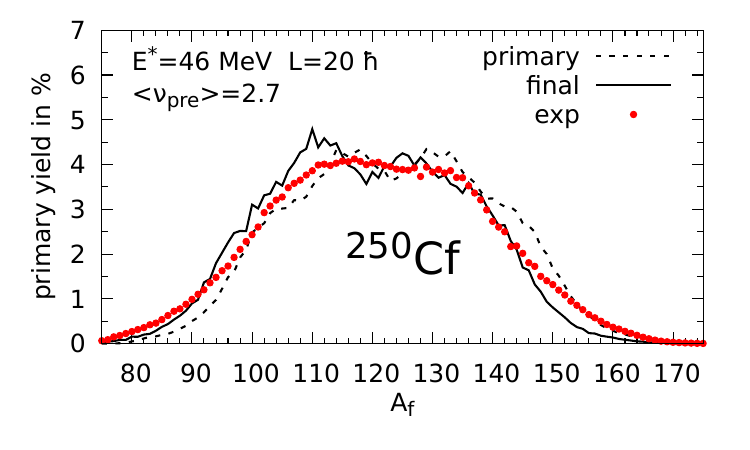}
\caption{Primary (dashed line) and final (solid line) fission fragment mass
 yields of $^{250}$Cf obtained without (top) and with (bottom), considering
 multi-chance fission. The experimental data (red diamonds) are taken from
 Ref.~\cite{RCF19}.}
\label{Cf250mexp}
\end{figure}

\begin{figure}[h!]
\includegraphics[width=0.45\textwidth]{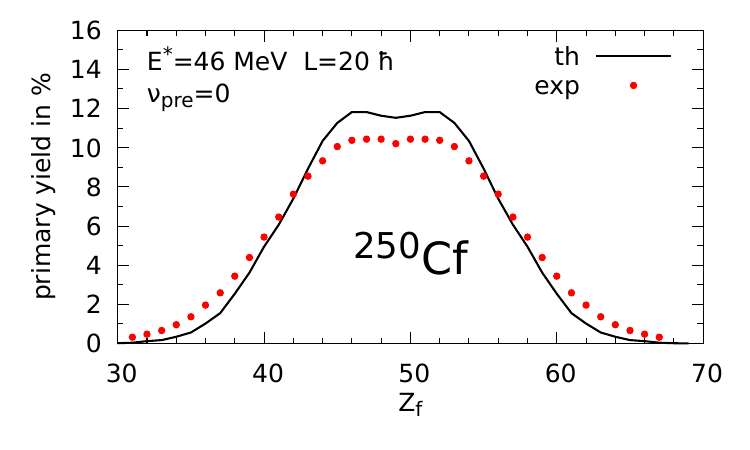}\\
\includegraphics[width=0.45\textwidth]{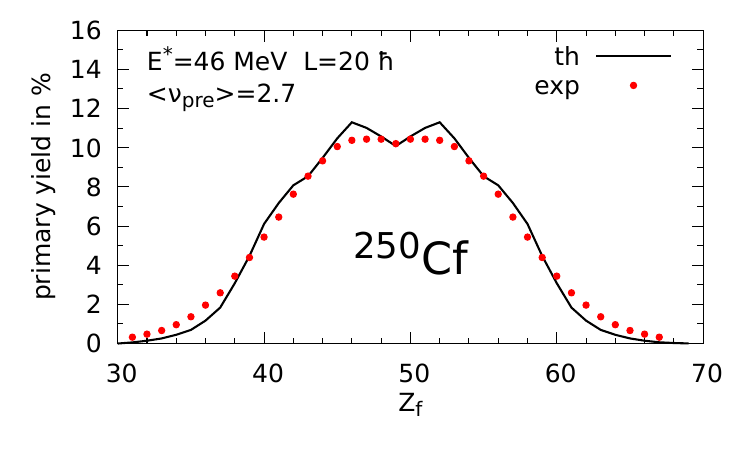}
\caption{Similar to Fig.~\ref{Cf250mexp} but for the charge yields.}
\label{Cf250chexp}
\end{figure}

As has been shown in Ref.~\cite{KDN21}, the fission fragment yields are to a good approximation independent of the initial conditions when the Langevin trajectories are started in the region of the scission point or at a smaller elongation of the fissioning nucleus. To allow for multi-chance fission but keep the computing time within reasonable limits, we have therefore performed five independent Langevin calculations for $^{246-250}$Cf isotopes with the initial thermal excitation energies as listed in Table I, starting from such an elongated initial configuration. Qualitatively, the PES's of these less excited $^{246-250}$Cf isotopes are intermediate between Figs.~\ref{Cf252pes} and \ref{Cf250pes24}. The theoretical mass (top), charge (middle), and TKE (bottom) yields obtained for the different numbers of pre-fission neutrons are shown in Fig.~\ref{Cf250yn}. The yields obtained for each pre-scission isotope is then weighted with its probability (2nd raw in Table I). The calculated primary (without taking neutron evaporation into account) and secondary (including neutron evaporation) mass yields are compared in Fig.~\ref{Cf250mexp} with the experimental data taken from Ref.~\cite{RCF19}. Similar plots for the charge yields are presented in Fig.~\ref{Cf250chexp}.

\begin{figure}[htb]
\includegraphics[width=0.5\textwidth]{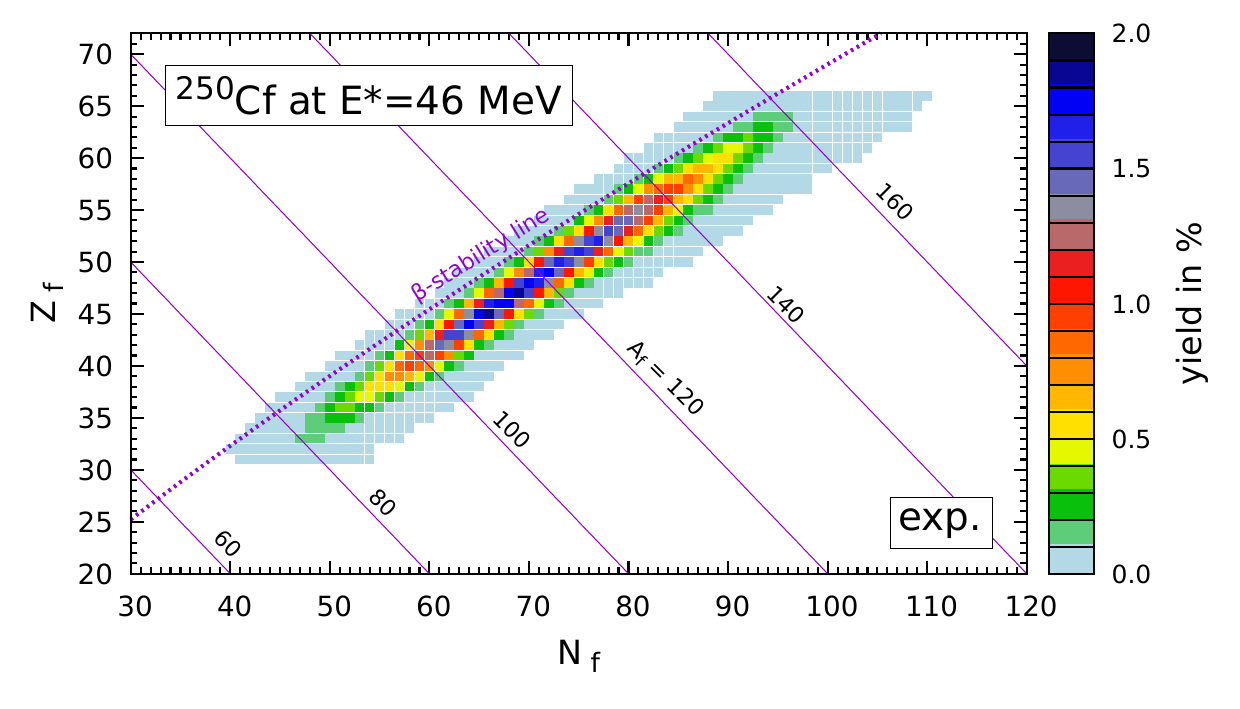}\\
\includegraphics[width=0.5\textwidth]{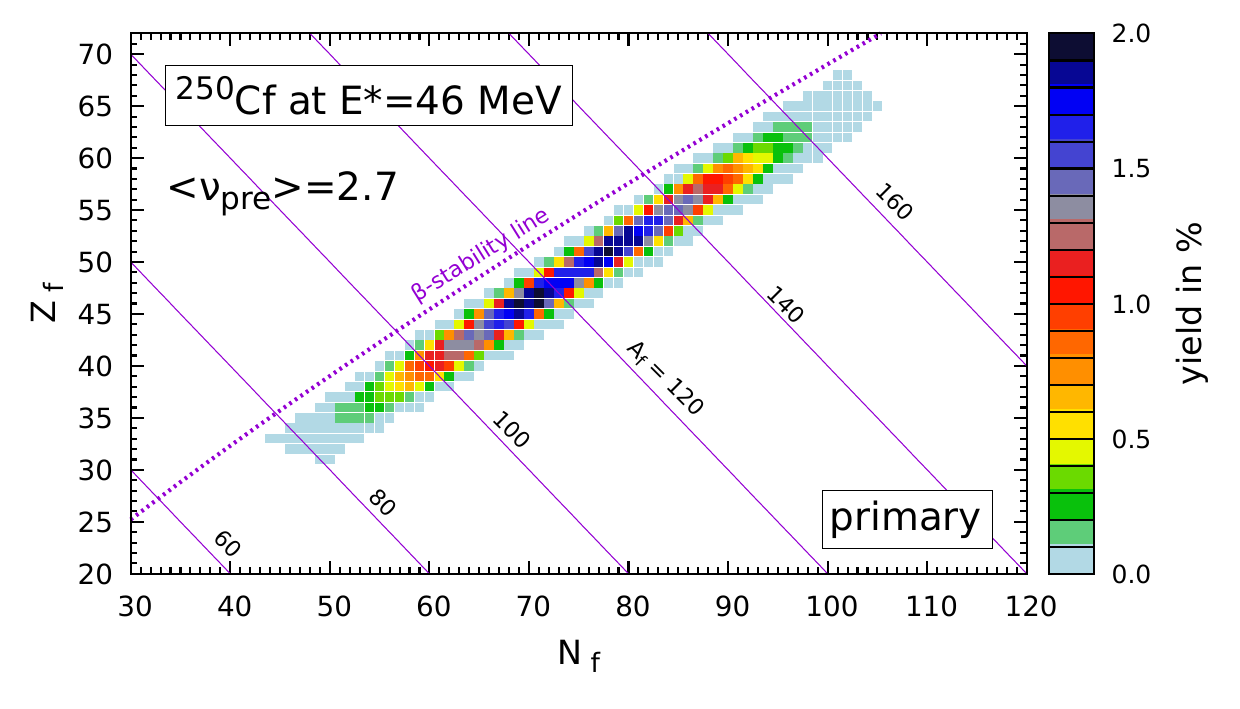}\\
\includegraphics[width=0.5\textwidth]{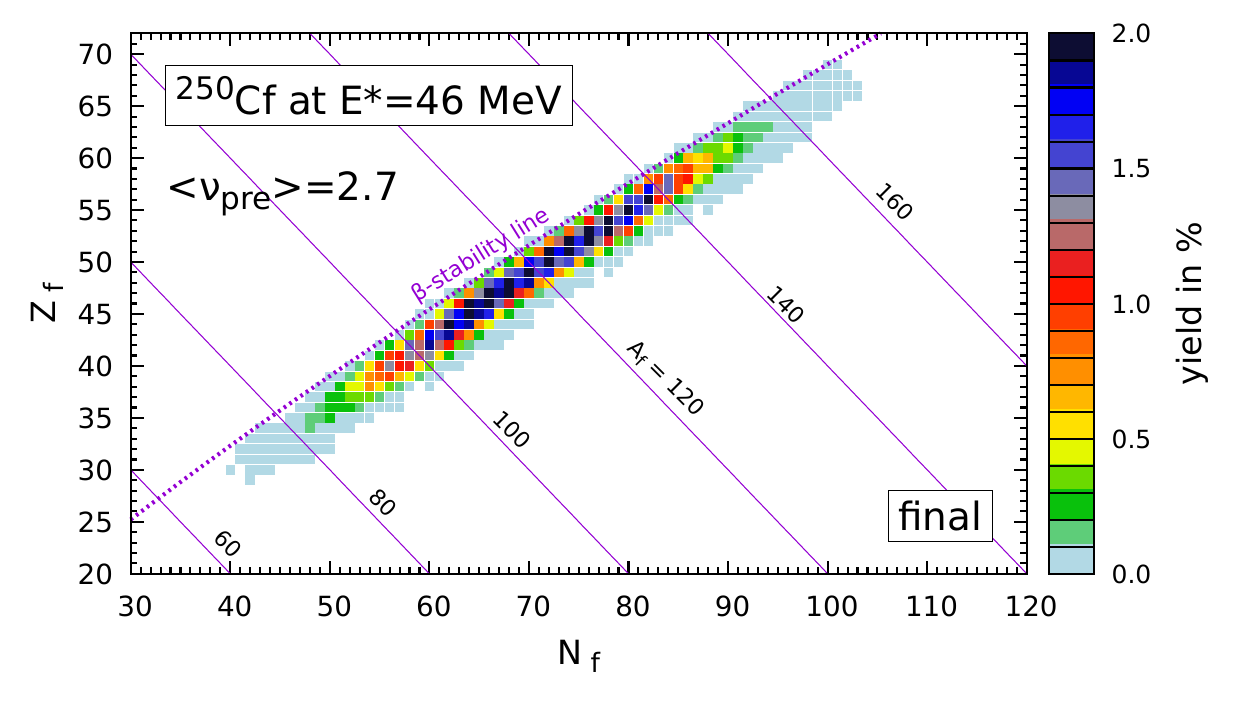}
\caption{Experimental (top), primary (middle), and final (bottom)
 isotopic yields of $^{250}$Cf. The experimental data are taken
 from Ref.~\cite{RCF19}.}
\label{Cf250nzy}
\end{figure}

It is seen in Figs.~\ref{Cf250mexp} and \ref{Cf250chexp} that the estimates obtained by taking into account the pre-fission neutron evaporation,  evaluated separately for different Cf isotopes and then weighted, are much closer to the data. The experimental (top), primary (middle), and final (bottom) estimates of the isotopic yields are shown in Fig.~\ref{Cf250nzy}  as functions of N$_{\rm f}$ and Z$_{\rm f}$. The calculations were based on $5\times 100\,000$ Langevin trajectories, so the range of less-probable nuclides is slightly smaller than the one obtained experimentally in Ref.~\cite{RCF19}. The final distribution of yields, i.e., after neutron emission from the fragment, is found to be shifted by 2-3 units relative to the measured ones.  
\begin{figure}[htb]
\includegraphics[width=0.5\textwidth]{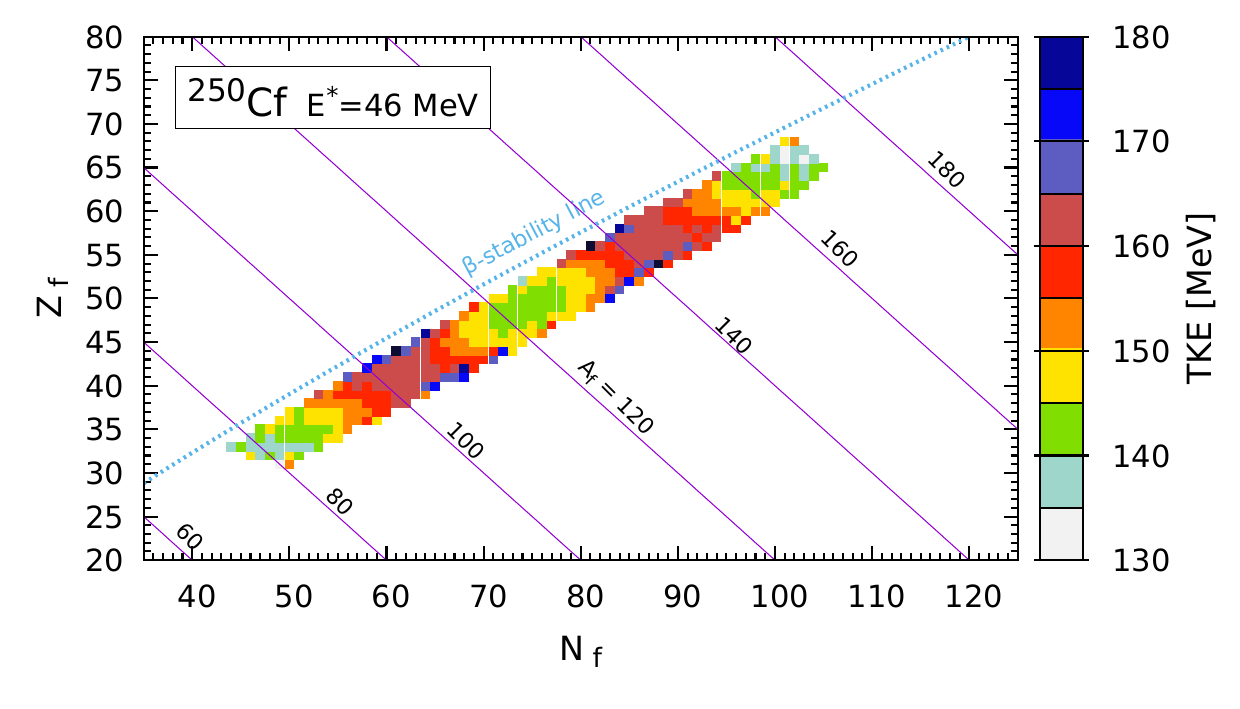}
\caption{Total kinetic energy (TKE) of as a function of $N_{\rm f}$ and $Z_{\rm f}$  primary fission fragment for $^{250}$Cf at $E^*$=46 MeV.}
\label{Cf250TKE}
\end{figure}
\begin{figure*}[t!]
\includegraphics[width=\textwidth]{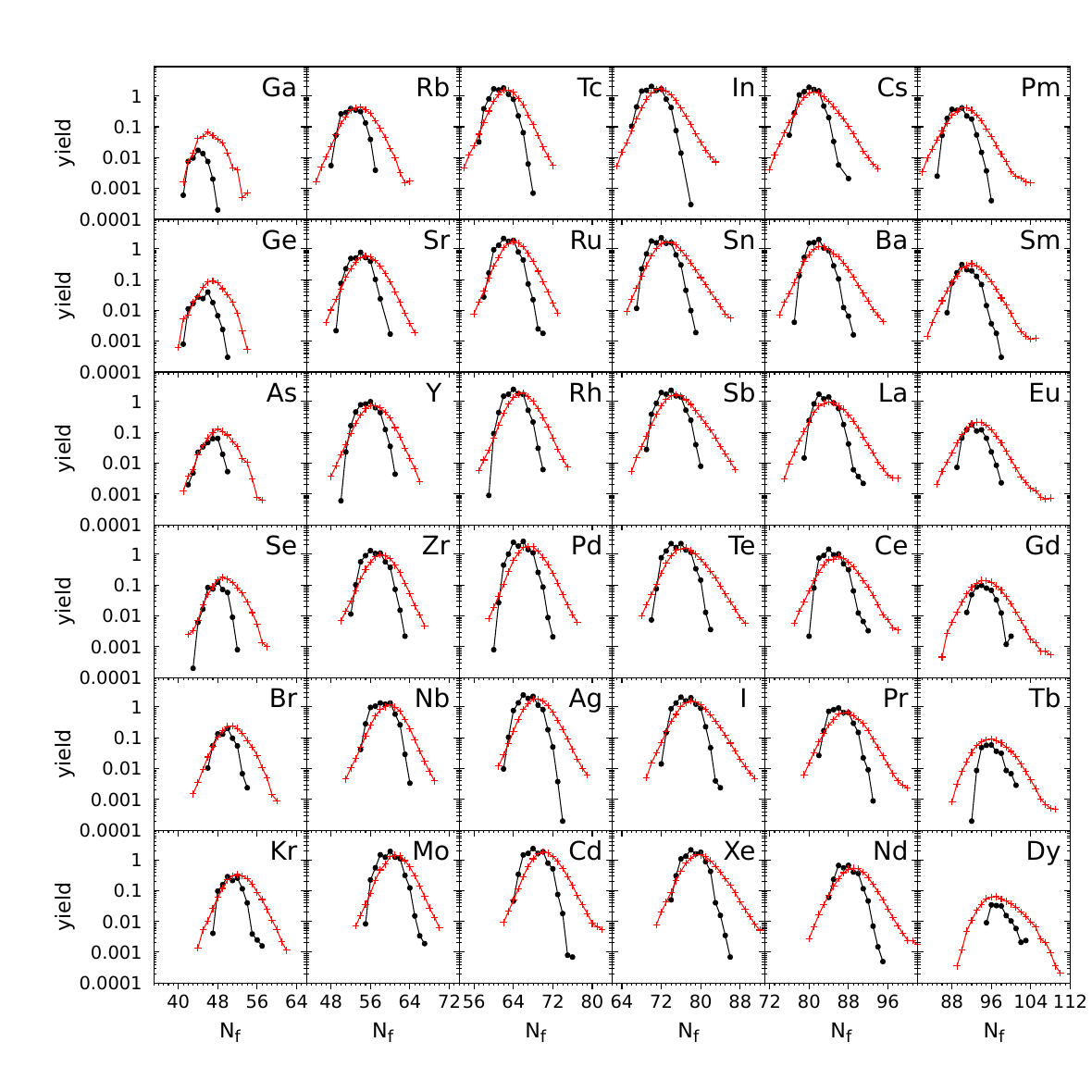}\\[-4ex]
\caption{Secondary fragment isotopic yields from Ga to Dy for fission of $^{250}$Cf at E*~46 MeV. Theoretical estimates are shown as black dots, and experimental data (red +) are taken from Ref.~\cite{RCF19}.}
\label{Cf250imy}
\end{figure*}

A similar plot but for the fragments' total kinetic energy (TKE) is shown in Fig.~\ref{Cf250TKE}. For the lightest and the heaviest fragments, as well as the ones corresponding to the symmetric fission, our model predicts a small TKE around 140 MeV, while the fragments with masses around $A=140$ or $A=110$ are found to have larger TKE's around 160 MeV.

A more detailed comparison of our model with the data \cite{RCF19} is shown in Fig.~\ref{Cf250imy} with the secondary isotopic distributions of fragment elements from Ga to Dy plotted as a function of the neutron number. Both theoretical and experimental yields show a kind of inverted parabola in the logarithmic scale. However, the stiffness of all the experimental distributions is significantly smaller than the ones of our theoretical estimates. In addition, the peak of the experimental distribution is generally shifted by 2-3 units towards larger neutron numbers relative to the theoretical distribution, as already deduced above. It is interesting to note that although the description of the integral mass and charge yields are of similar quality for spontaneous fission of $^{252}$Cf and fusion-fission of $^{250}$Cf, the predictions for the isotopic yields are slightly worse in the latter case. This effect may suggest some deficiency in treating multi-chance fission and/or evaporation in general. However, this observation still needs further investigation due to the interplay of various aspects during the fission process and the interdependence of its different stages.
\begin{figure}[htb]
\includegraphics[width=0.5\textwidth]{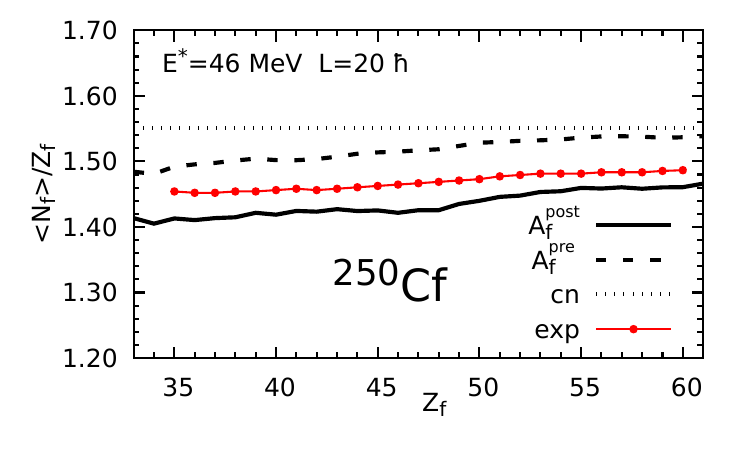}
\caption{The ${\rm N_{\rm f}/Z_{\rm f}}$  ratio of the fragment average number of neutrons
to the charge number for $^{250}$Cf at $E^*=$ 46 MeV. The solid line corresponds
to the post-neutron emission numbers, while the dashed one to the ratio
evaluated before neutron emission from the fragment. The dotted line marks the
neutron to proton ratio for $^{250}$Cf. The experimental data (red diamonds) are
taken from Ref. ~\cite{RCF19}.}
\label{Cf250NoZ}
\end{figure}
%om each fragment, we see that our model overestimates the number of neutrons emitted
\begin{figure}[t!]
\includegraphics[width=0.5\textwidth]{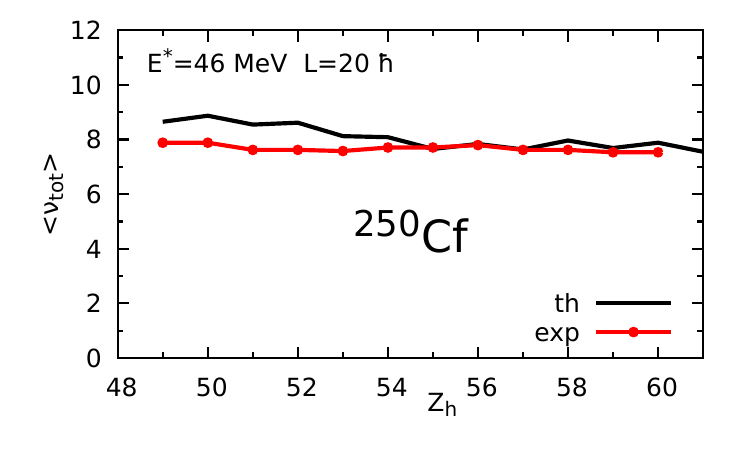}
\includegraphics[width=0.5\textwidth]{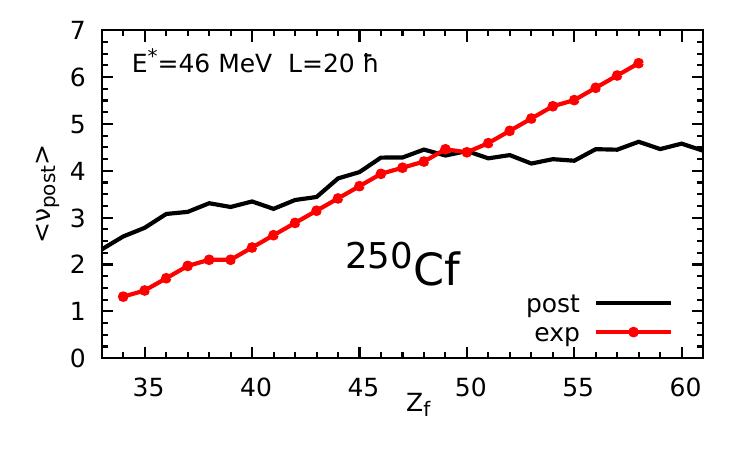}
\caption{The total number of neutrons emitted from both fission fragments for $^{250}$Cf at $E^*=$ 46 MeV (top) and the number of the neutrons emitted per fragment (bottom) as a function of the fragment charge number. The experimental data are taken from Refs.~\cite{RCF19} and \cite{CDF13}.}
\label{Cf250nu}
\end{figure}    

The ${\rm N_{\rm f}/Z_{\rm f}}$ ratio as a function of the fragment charge number is shown for $^{250}$Cf at $E^*=$ 46 MeV in Fig.~\ref{Cf250NoZ}. Our estimates corresponding to the primary (dashed line) and final (solid line) yields are compared with the data (red diamonds) taken from Ref.~\cite{RCF19}. The dotted line indicates the neutron-to-proton ratio in the parent nucleus. The experimental data are located in between the pre- and post-emission lines, which bears a suspicion that we overestimate the neutron number emitted from the fragments. As one can see in Fig.~\ref{Cf250nu} (top), the calculated {\it total} number of neutrons emitted from both fragments is described in a rather satisfactory way.
On the contrary, looking at Fig.~\ref{Cf250nu} (bottom) shows that the model overestimates the number of neutrons emitted from the light and underestimates the ones from the heavy fragments. Therefore, The deficiency above seems to be connected to the neutron emission balance between the two fragments and thus may be attributed to the description of the sharing of the nucleons and/or excitation energy at scission. However, due to the entangled process, further investigations are required before a final conclusion can be drawn.  
%
%%%%%%%%%%%%%%%%%%%%%%%%%%%%%%%%%%%%%%%%%%%%%%%%%%%%%%%%%%%%%%%%%%%%%%%%%%%%%%%
%%%%%%%%%%%%%%%%%%%%%%%%%%%%%%%%%%%%%%%%%%%%%%%%%%%%%%%%%%%%%%%%%%%%%%%%%%%%%%%
\section{Summary and conclusions}

In a previous communication \cite{PNS23} we presented a multi-dimensional Langevin fission model capable of handling the various facets of the process, including {\it i}) the dynamical evolution of the fissioning system between the ground state and the scission point, in competition with the particle evaporation,  {\it ii}) the sharing of neutrons, protons, and excitation energy between the two fragments at the moment of scission,  {\it iii}) their kinetic energy after full acceleration, and finally  {\it iv}) their decay back to equilibrium through the evaporation of neutrons.  The energy dependence of the different ingredients has been included from the beginning. The model was tuned and tested till now for low-energy fission only, particularly for thermal neutron-induced fission of $^{235}$U.  It also attested its capacity \cite{PNS23} to give a fair description of the evolution of the fragment properties along the Fermium isotopic chain in the low-energy regime where most experimental information is available.

In the present study, the theoretical framework developed in Ref.~\cite{PNS23} was applied, without any change of parameters, to the spontaneous fission of $^{252}$Cf and the fission of $^{250}$Cf produced at an excitation energy of 46 MeV in a fusion reaction,  thus permitting to test the predictive power of our model over an extended range of temperature, and thereby the implemented energy dependences.

A further extension of the present work compared to Ref.~\cite{PNS23} is the investigation of more detailed observables, particularly fragment isotopic distributions with unique resolution. The recent availability of such accurate data makes it possible to test fission models less ambiguously since previous data often needed better resolution or were restricted to integral distributions. Wherever the corresponding data are available, the model is found to describe reasonably well the integral primary and secondary mass and charge yields, the distribution of the fragment total kinetic energy, as well as the total amount of neutrons emitted in coincidence with fission for both $^{252}$Cf and $^{250}$Cf. The quite accurate reproduction of the isotopic yields for fragment elements from Ga to Dy shows a good description for spontaneous fission of $^{252}$Cf, but a somewhat poorer performance for higher excitation energy fission of $^{250}$Cf. The simultaneous analysis based on the total and individual ({\it viz.} per fragment) neutron multiplicities suggests a deficiency due to the properties of the fragments emerging at scission and probably with the calculated excitation energies. Further studies in this direction and other alternative explanations, such as e.g. charge equilibration and shell effects, will be the subject of future investigations.

The present study demonstrates the importance of accurate and high-fold correlation experimental information for constraining fission models. The availability of more and more data of this kind will be very beneficial to improve the present model, and fission theory in general. \\[4ex]

{\bf Acknowledgments}\\

The authors would like to thank D. Ramos, I. Mardor, and Y. Kehat for valuable discussions and for supplying us with experimental data. This work has been supported by the Polish-French agreement COPIN-IN2P3, project No. 08-131, and by the Natural Science Foundation of China (Grant No. 11961131010 and 12275081).\\[4ex]

\end{document}